\documentclass[doublecol]{epl2} 
\usepackage{mcite}
\usepackage{amssymb}

\title{Surface growth on treelike lattices and the upper critical dimension of the KPZ class}
\shorttitle{Surface growth on treelike lattices} %Insert here a short version of the title if it exceeds 70 characters

\author{Tiago J. Oliveira\thanks{E-mail: \email{tiago@ufv.br}}}
\shortauthor{T. J. Oliveira}

\institute{                    
  Departamento de F\'isica, Universidade Federal de Vi\c cosa, 36570-900, Vi\c cosa, MG, Brazil
}
\pacs{81.15.Aa}{Theory and models of film growth}
\pacs{05.40.-a}{Fluctuation phenomena, random processes, noise, and Brownian motion}
\pacs{89.75.Da}{Systems obeying scaling laws}

\abstract{
Aiming to investigate the upper critical dimension, $d_u$, of the KPZ class, in [EPL \textbf{103} (2013) 10005] some growth models were numerically analyzed using Cayley trees (CTs) as substrates, as a way to access their behavior in the infinite-dimensional limit, and some unexpected results were reported: logarithmic roughness scaling, differing for EW and KPZ models (indicating that even at $d=\infty$ the KPZ nonlinearity is still relevant); beyond asymptotically rough EW surfaces above the upper critical dimension of the EW class. Motivated by these strange findings, I revisit these growth models here to show that such results are simple consequences of boundary effects, inherent to systems defined on CTs. In fact, I demonstrate that the anomalous boundary of the CT leads the growing surfaces to develop curved shapes, which explains the strange behaviors previously found for these systems, once the global ``roughness'' were analyzed for non-flat surfaces in the study above. Importantly, by measuring the height fluctuations at the central site of the CT, which can be seen as an approximation for the Bethe lattice, smooth surfaces are found for both EW and KPZ classes, consistently with the behavior expected for growing systems in dimensions $d \geqslant d_u$. Interesting features of the 1-pt height fluctuations, such as the possibility of non-saturation in the steady state regime, are also discussed for substrates in general.
}

\begin{document}

\maketitle

The stochastic partial differential equation by Kardar-Parisi-Zhang (KPZ) \cite{KPZ}
\begin{equation}
 \frac{\partial h(\vec{x},t)}{\partial t} = \nu \nabla^2 h + \frac{\lambda}{2} (\nabla h)^2 + \sqrt{D}\eta(\vec{x},t)
\label{eqKPZ}
\end{equation}
where $\eta$ is a Gaussian white-noise, describes the nonequilibrium dynamics of a vast number of physical systems, which are said to belong to the KPZ universality class. One of its most relevant applications is in growth phenomena, where $h(\vec{x},t)$ represents the height field of evolving interfaces \cite{barabasi,KrugAdv}. However, it may represent also the energy of polymers in random media \cite{healy95}, the density of particles in one-dimensional (1D) transport processes \cite{Kriecherbauer2010} and so on. In all cases, $\nu$, $\lambda$ and $D$ are phenomenological parameters. Moreover, when $\lambda=0$ one recovers the Edwards-Wilkinson (EW) equation \cite{EW}.

The 1D KPZ class is quite well-known in several respects. For instance, the exponents for the scaling of the (global) surface roughness, $W$, as a function of time ($W \sim t^{\beta}$ --- during the transient growth regime) and with the system size $L$ ($W_s \sim L^{\alpha}$ --- at the steady state regime) are exactly known as $\beta=1/3$ and $\alpha=1/2$ \cite{KPZ}.  This last exponent is the same as for the EW class, as a consequence of the irrelevance of the KPZ nonlinearity in the steady state regime in the 1D case \cite{KPZ}. This, by the way, means that the probability density function (pdf) for the height fluctuations --- i.e., the underlying height distribution (HD) --- is Gaussian in this regime. The asymptotic KPZ HDs are also known for the growth regime, where they are given by Tracy-Widom \cite{TW} distributions from different ensembles, depending on the surface geometry \cite{Prahofer2000,*Sasamoto2010,*Amir,*Calabrese2011,*Imamura}. This geometry-dependence has motivated a large number of studies in recent years, confirming the universality of HDs \cite{Takeuchi2010,*Takeuchi2011,*Alves11,*tiago12a,*HealyCross,*Silvia17,*Sidiney,*Rodolfo2020,Alves13}, as well as generalizing this for other universality classes \cite{Ismael16a,*Ismael19b,*Ismael20} and other situations within the 1D KPZ class \cite{Fukai2017,*Ismael18,SaberiKrug,Ismael19}.

In higher dimensions the KPZ picture is far less understood. Despite the scaling relation $\alpha + \alpha/\beta = 2$, due to Galilean invariance \cite{barabasi}, it seems that no more rigorous results are available. For instance, in the 2D KPZ class (the most important for surface growth) the exponents calculated from different analytical approaches \cite{Lassig,Moore,Fogedby} present discrepancies with the more accurate numerical estimates for them \cite{Kelling,Pagnani} (see, e.g., Ref. \cite{Pagnani} for a detailed discussion on this). Moreover, although the HDs have been accurately determined for a number of 2D KPZ models and display also a dependence with geometry \cite{healy12,*healy13,*Ismael14,tiago13}, their exact pdf's are not known yet.

Another long-standing and controversial issue in the KPZ class is its upper critical spatial dimension, $d_u$, at which the power-law behaviors of the roughness $W$ give place to logarithmic scaling and above which KPZ surfaces shall be smooth \cite{barabasi}. In fact, while most of the studies based on mode-coupling theory \cite{Moore,Cates,*Bray,*Claudin} and field theoretical approaches \cite{Healy90,*Kinzelbach,*Fogedby2} have predicted $2.8 \leq d_u \leq 4$, renormalization group calculations indicate that $d_u$ is larger than four \cite{Canet,*Kloss,Perlsman0,*Castellano1,*Castellano2}, with some of them pointing at $d_u = \infty$ \cite{Perlsman0,*Castellano1,*Castellano2}. From a numerical side, strong evidence also exists that, if $d_u$ is finite, it is larger than six \cite{Ala1,*Ala2,*Marinari02,*Perlsman2,*Perlsman1,*Geza10,*Parisi13,*JMKim13,*Fernando,*Alves16,*JMKim,*Alves14}. 

In light of this, Saberi \cite{Saberi} had the clever idea of numerically investigating EW and KPZ growth models using a Bethe lattice (BL) as substrate, as a way to compare their behavior in the infinite-dimensional limit. As an aside, I recall that $\alpha=(2-d)/2$ and $\beta=(2-d)/4$ for the EW class, with $d$ being the substrate dimension, so that $d_u^{EW}=2$. Since different scaling behaviors were numerically found for these classes, it was claimed in \cite{Saberi} that the KPZ nonlinearity is still relevant even for $d \rightarrow \infty$, \textit{``consequently questioning the existence of a finite upper critical dimension for the KPZ class''}. It turns out however that if the KPZ systems are in the strong-coupling regime even in infinite dimension, this completely rules out the possibility of an upper critical dimension. Namely, $d_u$ should neither be finite nor infinite, which seems to be quite unexpected. Another very strange result reported in \cite{Saberi} was the rough surfaces found for the EW class, with $W_s \sim k^{0.6}$ in the steady state regime, where $k$ is the number of generations of the tree considered as substrate. In fact, this result would imply that $W_s \rightarrow \infty$ when the substrate size goes to infinity ($k \rightarrow \infty$), for an infinite-dimensional EW system, whose surfaces should be smooth (i.e., $W_s \sim const.$ and small) already for $d \geqslant 3$. Given the theoretical relevance of this problem, it is quite important to further examine growth models on the BL in order to deeply understand the issues raised in \cite{Saberi}. As will be demonstrated in what follows, these results are artifacts due to the boundary of the substrate considered, allied to a wrong calculation of the surface roughness. In other words, Saberi \cite{Saberi} has investigated the models on finite Cayley trees (CTs), which are dominated by an anomalous boundary, rather than on BLs, where boundary effects are absent. By considering the correct approximation for the BL (i.e., by analyzing the 1-point height fluctuations, $w_0$, at the central site of the CT), I find here similar fluctuations for the EW and KPZ models, which are consistent with the behavior expected for these growing systems at $d = \infty$. By the way, the scaling properties of $w_0$ are studied in detail, for substrates in general, and key differences from the global roughness $W$ are revealed.

I investigate, through extensive Monte Carlo simulations, the same models considered in \cite{Saberi}. In all cases, particles are randomly deposited on a CT of $N_T$ sites, which is initially flat (i.e., $h_j = 0$ for $j=1, \ldots, N_T$). In the restricted solid-on-solid (RSOS) model \cite{kk}, aggregation is accepted at site $i$ only if it yields $|h_i - h_j|\leq 1$, with $j$ representing all nearest neighbors (NNs) of $i$; otherwise, the particle is rejected. In the ballistic deposition (BD) model \cite{barabasi}, the particles aggregate at their first contact with the surface. In the Family model \cite{Family}, aka random deposition with surface relaxation (RDSR) model, the freshly deposited particle can diffuse in a limited area on the surface (up to second neighbors here) in order to minimize its height. In all models, the time is increased by $1/N_T$ after each \textit{aggregation} event, so that in the Family and RSOS cases the spatially averaged height is given by $\bar{h}(t)=t$. Some results will be presented also for the commonest version of the RSOS model in literature (I will refer to it as the RSOSc model), which is almost identical to the previous version, the only difference being that the time is increased by $1/N_T$ after each deposition \textit{attempt}. In this case, as well as in BD model $\bar{h}(t)$ is a fluctuating variable. There is a variety of evidence in literature that, at least on regular substrates, the BD and the RSOS models belong to the KPZ class, while the RDSR model is in the EW class.

\begin{figure}[t]
\centering
 \includegraphics[width=7cm]{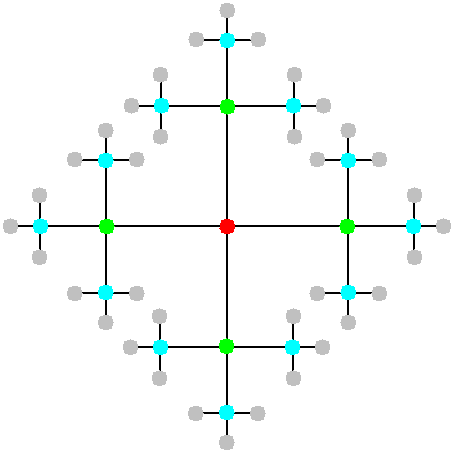}
 \caption{Symmetric Cayley tree with three generations ($k=3$) and coordination $q=4$. Different colors represent sites in different shells of the tree.}
 \label{fig1}
\end{figure}

Let me start recalling that a CT is a connected graph, without loops, where all vertices (or sites) have the same degree (or coordination) $q$, with exception of those at the boundary, which have degree $q=1$ (i.e., a single NN). A symmetric CT can be built by starting with a single site at the origin and connecting $q$ sites to it through $q$ edges, forming an 1-generation tree. Then, by successively adding $q-1$ new sites to each boundary site of the previous generation (or shell), a $k$-generation tree can be built, as seen in Fig. \ref{fig1}, for $q=4$ and $k=3$. The number of boundary sites of a CT with coordination $q$ and $k$ generations is $N_b = q (q-1)^{k-1}$, while in its interior there are $N_i = (N_b-2)/(q-2)$ sites, for $q>2$. Hence, this tree has a large fraction of boundary sites, even in the thermodynamic limit ($k\rightarrow \infty$), where $N_b/N_i = (q-2)$, in contrast to regular lattices, for which $N_b/N_i \rightarrow 0$ in this limit. 

As a consequence of this, models considered on the CT have their behaviors determined by boundary effects, which might not represent their correct physics on regular lattices. A classical example is the Ising model, which displays quite unusual properties on the CT \cite{Eggarter,*Hartmann,*Gandolfo}. On the other hand, in the core of an infinite CT the solution of the Ising model was found by Kurata \textit{et al.} \cite{Kurata} to be equivalent to the Bethe approximation \cite{Bethe}. Because of this, the \textit{core of the infinite CT} --- i.e., its central region, infinitely far from the boundary --- became known as the \textit{Bethe lattice} \cite{BaxterBook}. Concerning finite CTs, it seems that Woodbury Jr. \cite{Woodbury} and Runnels \cite{Runnels} were the first to investigate their equivalence with the BL, for the hard lattice gas model, finding that the Bethe approximation is not exact over the whole tree, albeit it can be recovered for some model parameters, by eliminating the boundary effects. Namely, considering only the central portion of the tree. The literature is plenty of works considering solutions of models on the BL and, in all cases, they are obtained in a way that the quantities of interest are calculated in the core of the infinite tree. Although it is not clear how this type of exact solution can be obtained for growth models, it seems reasonable to expect that their height fluctuations at the center of finite CTs shall be a good approximation for those in the BL. For further discussions on CT \textit{versus} BL, I invite the reader to see, e.g., Refs. \cite{BaxterBook,Gujrati,Ostilli}.

\begin{figure}[t]
 \includegraphics[width=4.25cm]{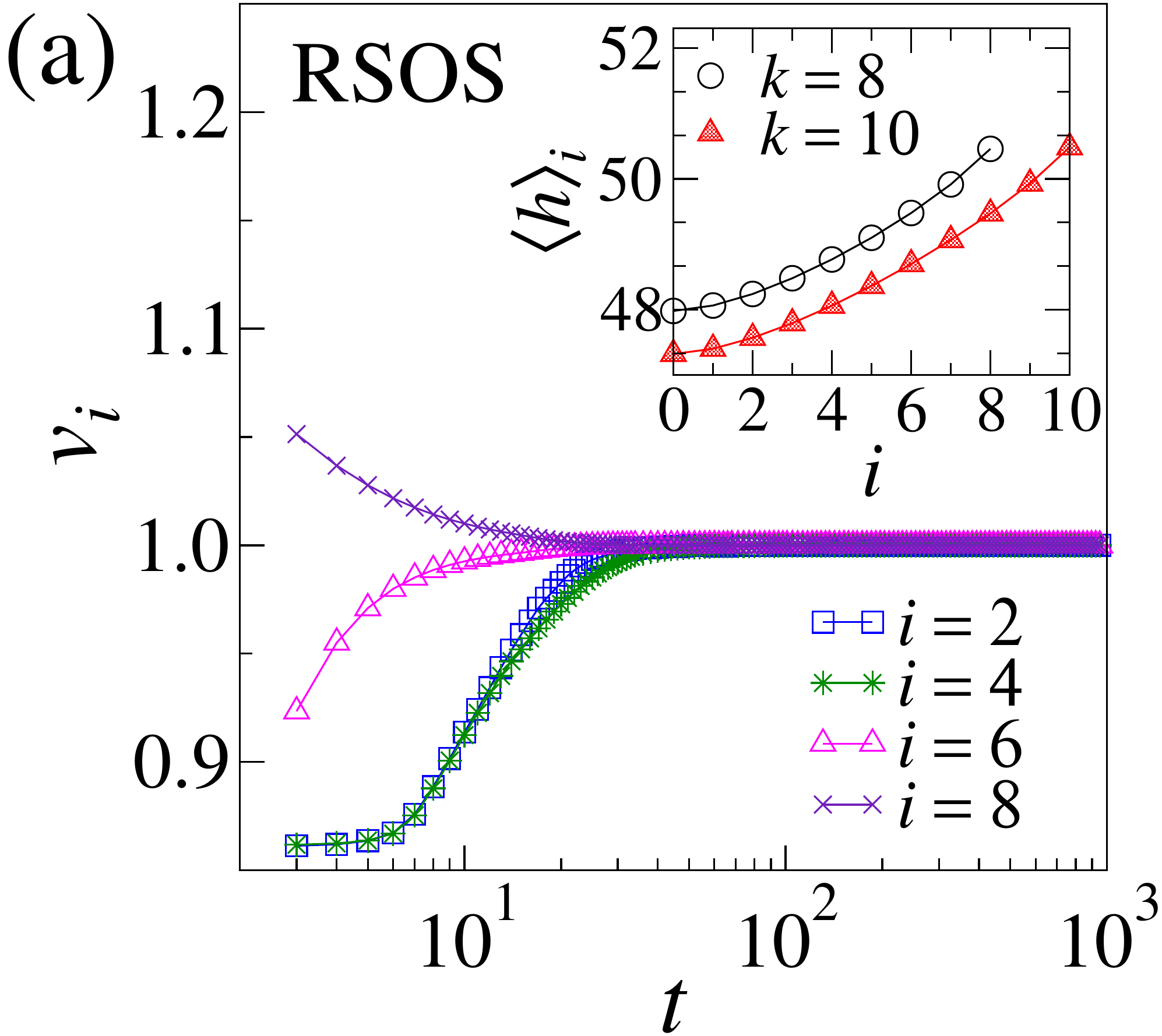}
 \includegraphics[width=4.25cm]{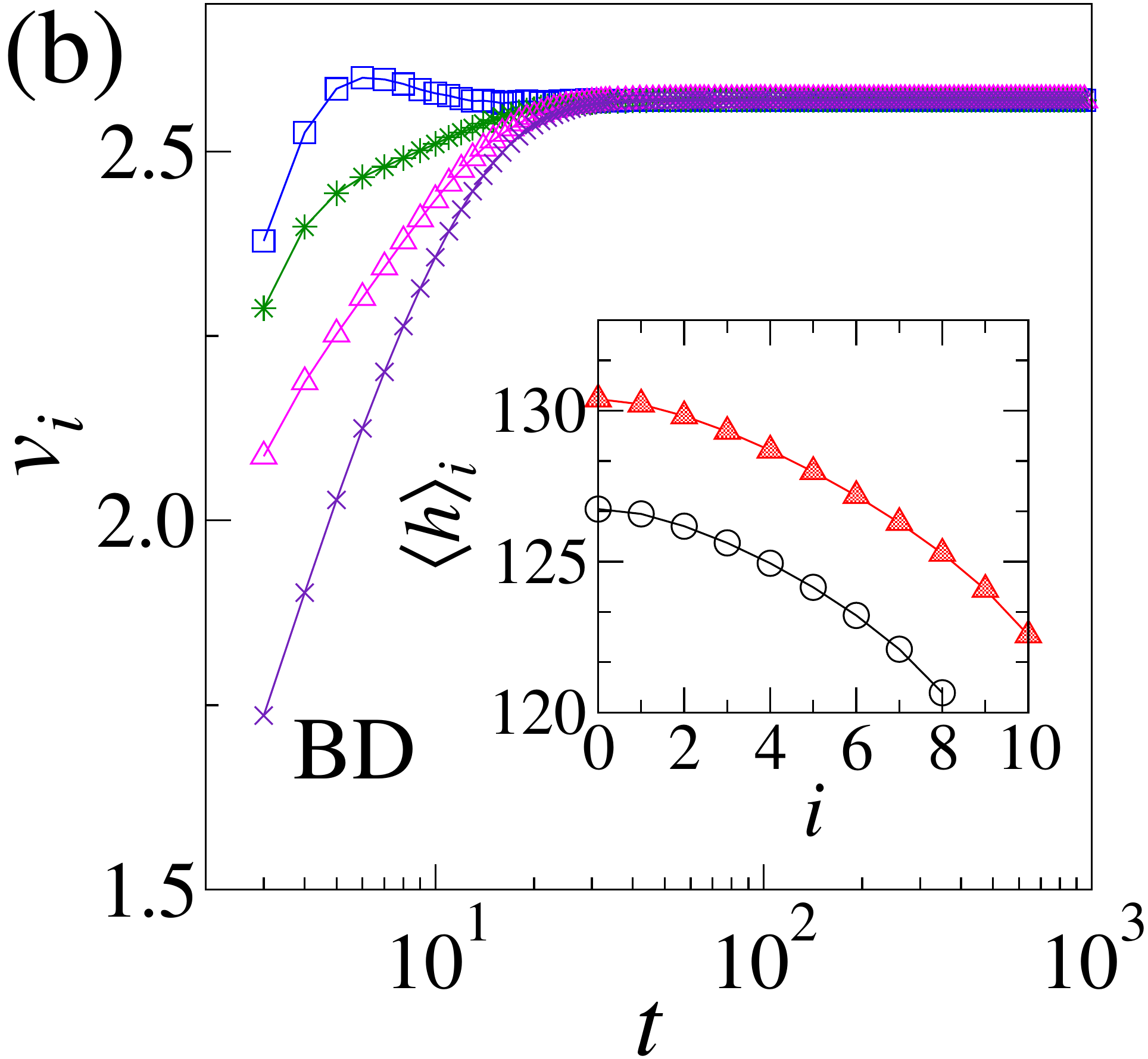}
 \includegraphics[width=4.25cm]{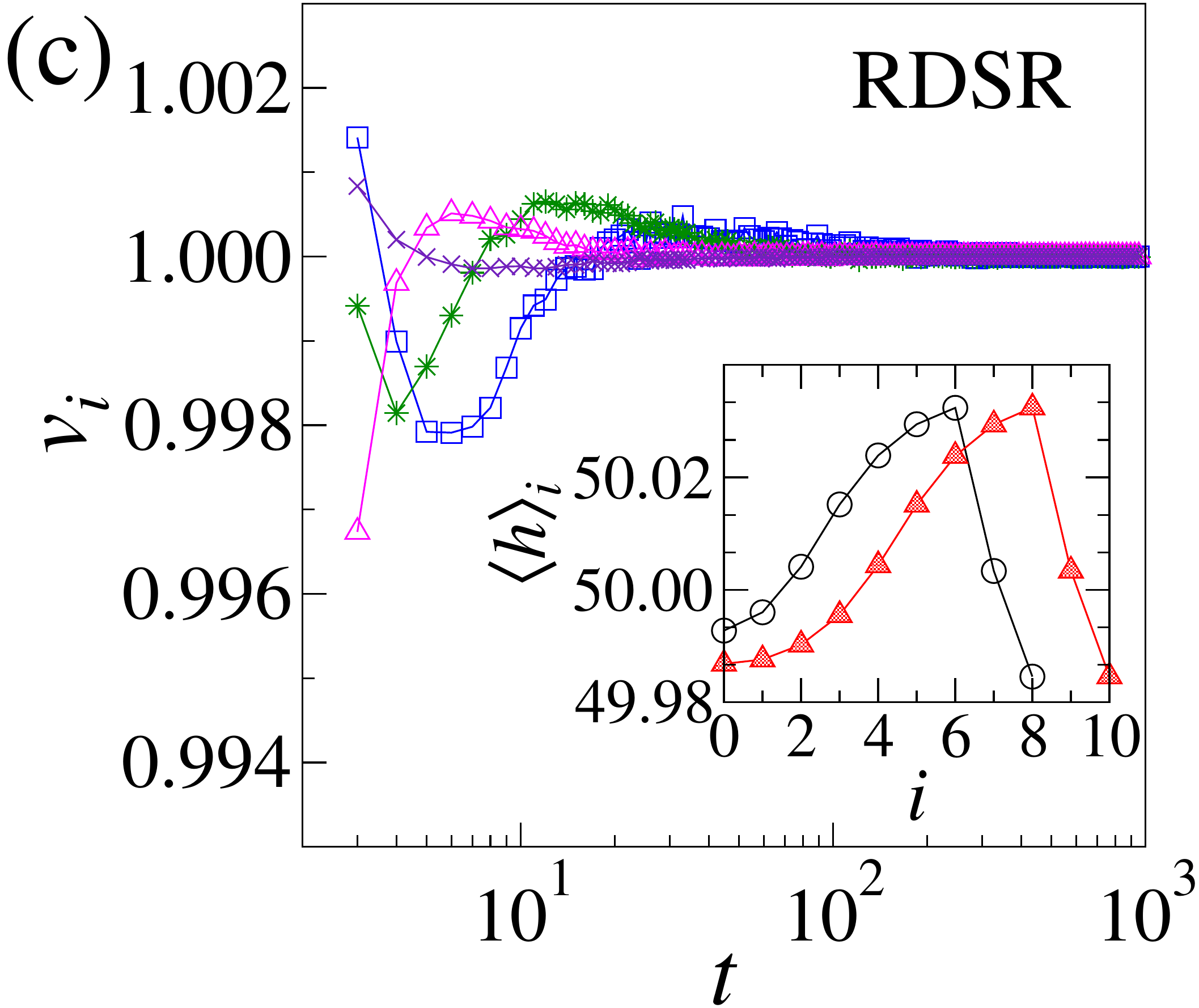}
 \includegraphics[width=4.25cm]{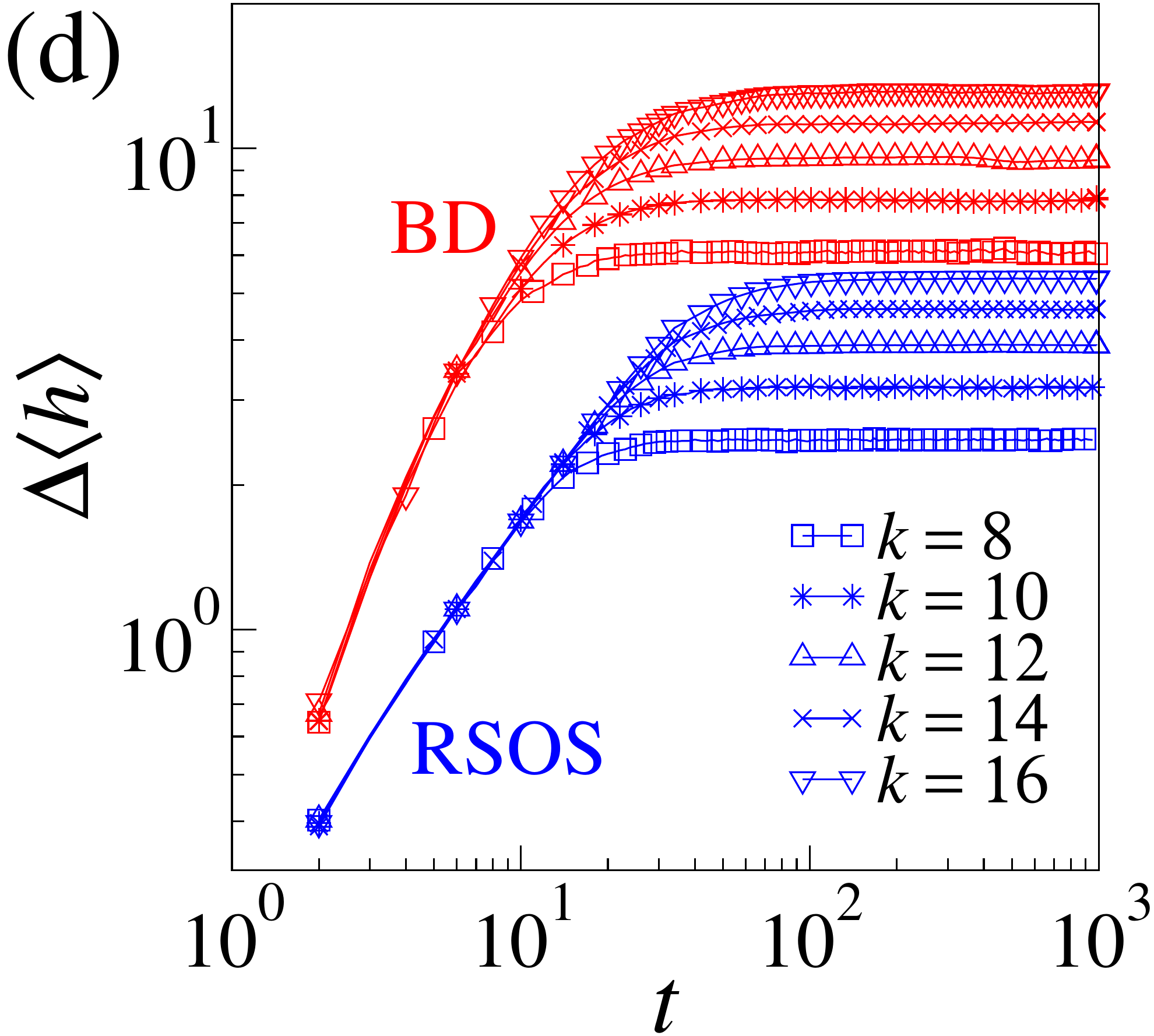}
 \includegraphics[width=4.25cm]{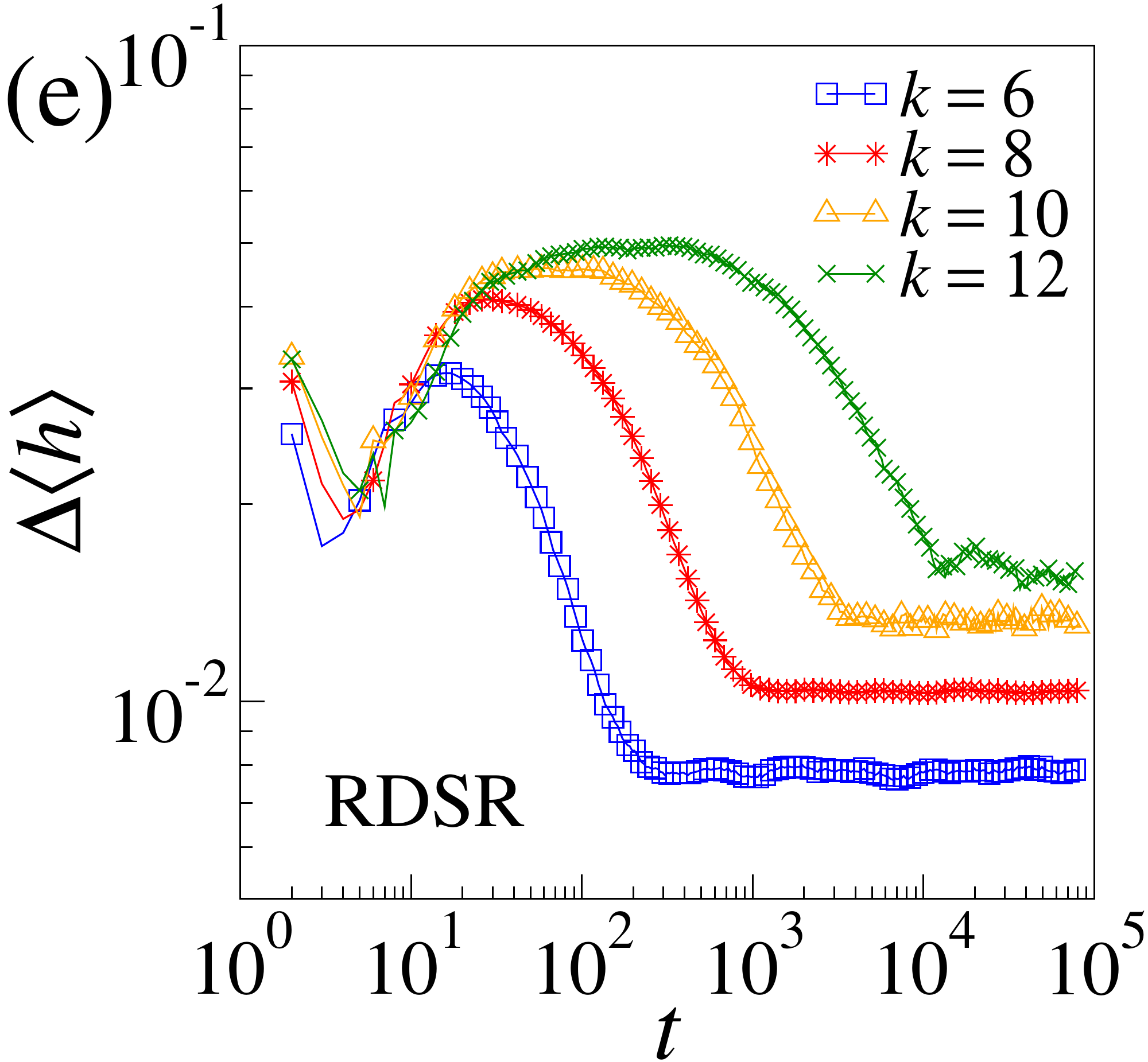}
 \includegraphics[width=4.25cm]{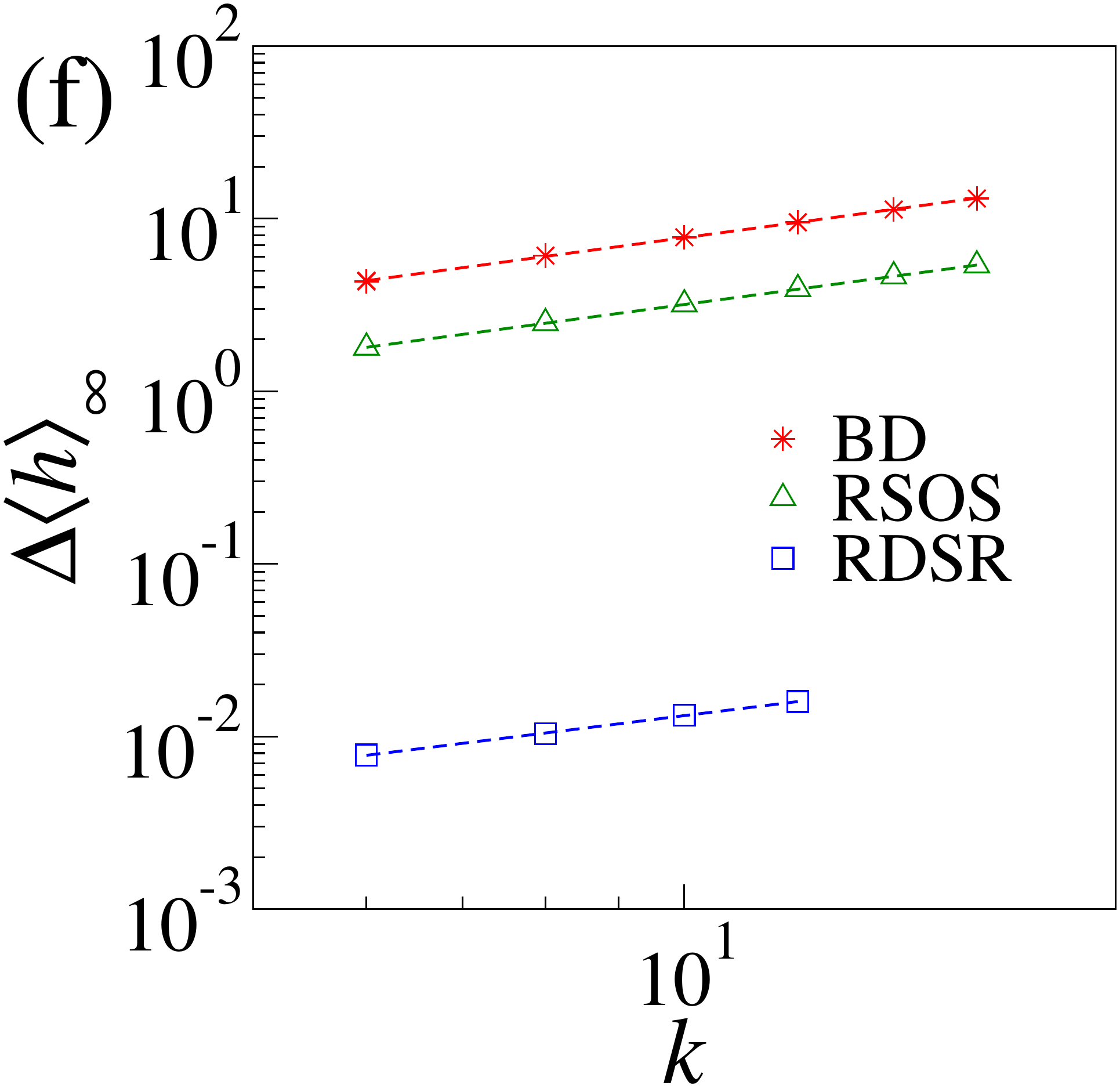}
 \caption{Growth velocity $v_i$ versus time $t$ [in linear-log scale] for the a) RSOS, b) BD and c) RDSR models simulated on CTs with $q=3$ and $k=8$. The insertions show the values of $\left\langle h \right\rangle_i$ at time $t=50$ against the shell $i$, where data for $k=8$ and $k=10$ are presented. Data in panels (b) and (c) are for the same parameters indicated by the legends in (a). The temporal evolution of the difference $\Delta \left\langle h \right\rangle$ is depicted in (d) for the KPZ models and (e) for the RDSR model, on CTs with $q=3$ and several $k$'s. In (f) the asymptotic values of this difference, $\Delta \left\langle h \right\rangle_{\infty}$, are shown as function of $k$. The dashed lines in (f) are linear fits of the data, while the solid lines in the other panels are guides to the eye.}
 \label{fig2}
\end{figure}

For growth models, it is simple to realize that the deposition dynamics at the boundary might be different from that in the interior of the CT, because the boundary sites have a single neighbor, while the others have $q$ NNs, being $q\geq 3$. So, in correlated growth models, this yields different aggregation rates at the center and boundary of the CT, at short times. Given the symmetry of the CT, these rates might be different for each shell $i\in[0,k]$. In fact, if one denotes the mean of the heights at shell $i$ by $\left\langle h \right\rangle_i$, its growth velocity $v_i=\partial \left\langle h \right\rangle_i/\partial t$ does indeed depend on $i$, as confirmed in Figs. \ref{fig2}a-c. These figures show examples of the temporal evolution of $v_i$, for the three models and several $i$'s, on CTs with $q=3$ and $k=8$. At long times, however, $v_i$ becomes independent of $i$, with $v_{\infty} = 1$ for the RSOS and RDSR models, and $v_{\infty} = 2.5696(3)$ in the BD case. Note that this is indeed expected, because otherwise the surface shapes would keep changing forever. Since the heights at different shells evolve with different velocities at short times, $\left\langle h \right\rangle_i$ becomes also a function of $i$. The insertions in Figs. \ref{fig2}a-c present $\left\langle h \right\rangle_i$ versus $i$, for $t=50$, which show the surface shapes developed in each model. In the RSOS case, $\left\langle h \right\rangle_i$ increases with $i$ because at short times it is easier to satisfy the RSOS condition for sites with a single neighbor at the boundary than for sites with $q\geq 3$ neighbors at the center of the tree. This compels the initially flat surface to convert into a curved one, as seen in Fig. \ref{fig2}a, in order to turn $v_i$ independent of $i$ at long times. This process is very similar to that recently discussed in Ref. \cite{SaberiKrug} for crossing KPZ interfaces. In the BD model, the higher coordinated sites at the CT's center increase faster than the lower coordinated ones at its boundary, at short times. Thereby, in this case $\left\langle h \right\rangle_i$ decreases with $i$, as observed in Fig. \ref{fig2}b. For the RDSR model, a more complex scenario is found in Fig. \ref{fig2}c, with $\left\langle h \right\rangle_i$ increasing with $i$ up to $i=k-2$, after which it has a fast decreasing. This certainly happens because in the RDSR model, as implemented here, the particles are allowed to diffuse up to second neighbors looking for the minimal height to aggregate there. Since particles deposited at the boundary have a single direction to diffuse, this yields a net current of them towards the shell $(k-2)$, explaining its larger mean height.

To gain some insight on how the surface shapes change in time, it is interesting to analyze the difference between their mean height at the center and periphery of the CT: $\Delta \left\langle h \right\rangle = |\left\langle h \right\rangle_0 - \left\langle h \right\rangle_j|$, with $j=k$ in the KPZ models and $j=k-2$ in the RDSR one. Figure \ref{fig2}d shows the evolution of this difference for the RSOS and BD models, where an increasing behavior is observed at short times, followed by a saturation of $\Delta \left\langle h \right\rangle$. For the RDSR model, an increase in $\Delta \left\langle h \right\rangle$ is also seen at short times, as depicted in Fig. \ref{fig2}e, but it is followed by a relaxation regime, where $\Delta \left\langle h \right\rangle$ decreases, before reaching its steady state. In all models, the asymptotic values of the difference, $\Delta \left\langle h \right\rangle_{\infty}$, increase with $k$ consistently with a power law scaling, $\Delta \left\langle h \right\rangle_{\infty} \sim k^{\gamma}$, as demonstrated in Fig. \ref{fig2}f. The linear fits in this figure return $\gamma \approx 1.12$ for both KPZ models and $\gamma \approx 1.03$ for the RDSR model. Such behaviors strongly suggest that the evolving surfaces will be macroscopically curved even for infinite CTs, for all models.

These results reveal a serious problem in the previous study of growth models on the CT \cite{Saberi}: the surface ``roughness'' (or ``width'') was calculated in \cite{Saberi} considering all surface points, as if they were flat. Namely, Saberi has measured the standard deviation of non-flat surfaces and wrongly reported this as the surface roughness. This is certainly the source of the strange results found in \cite{Saberi}. In fact, when the surfaces are non-flat (and non-spherical) the spatial translation symmetry is lost and the height fluctuations have to be measured for a single or a few (statistically equivalent) surface points \cite{Alves13,tiago13,SaberiKrug,Henkel}.

\begin{figure}[t]
 \includegraphics[width=4.25cm]{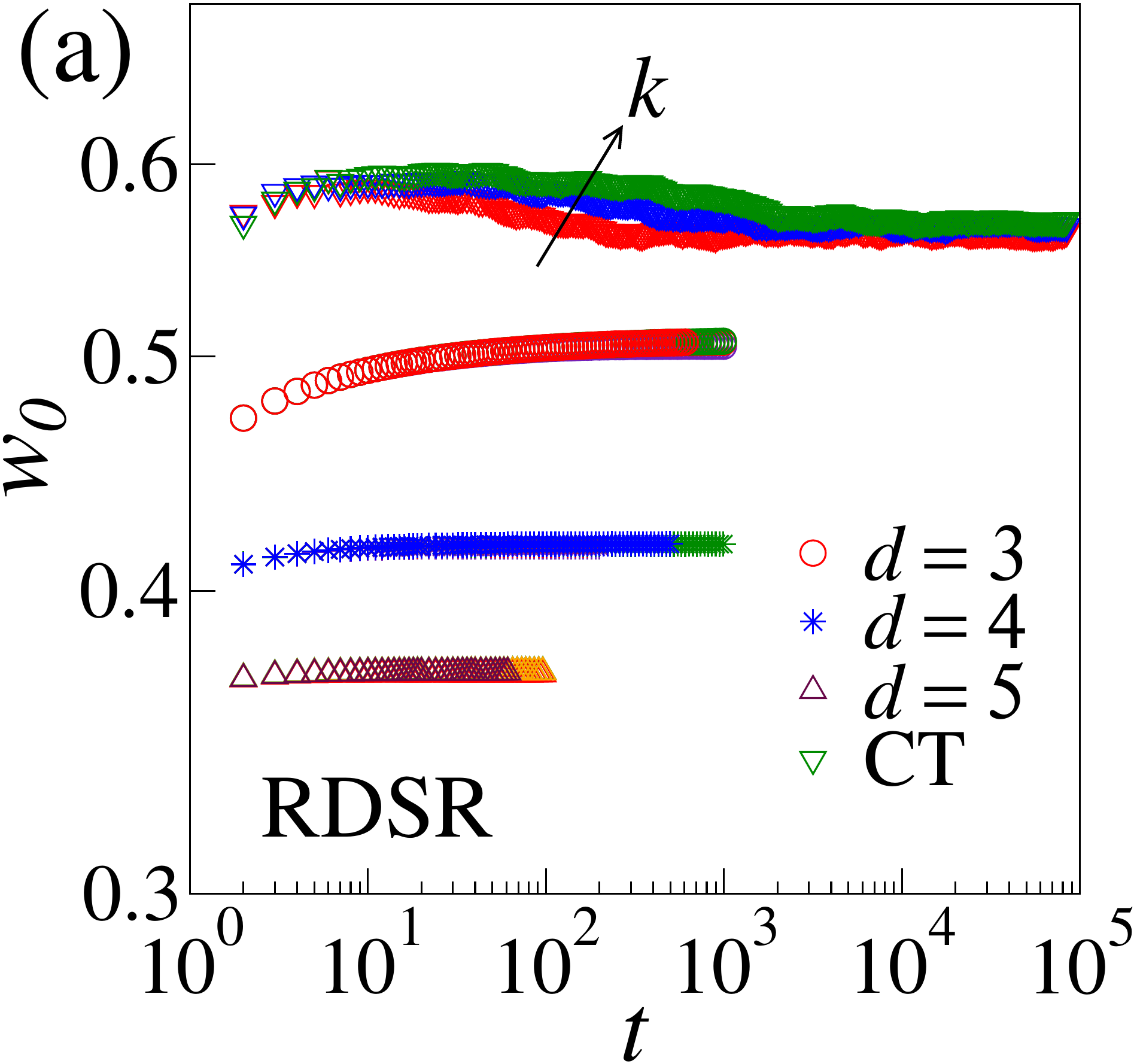}
 \includegraphics[width=4.25cm]{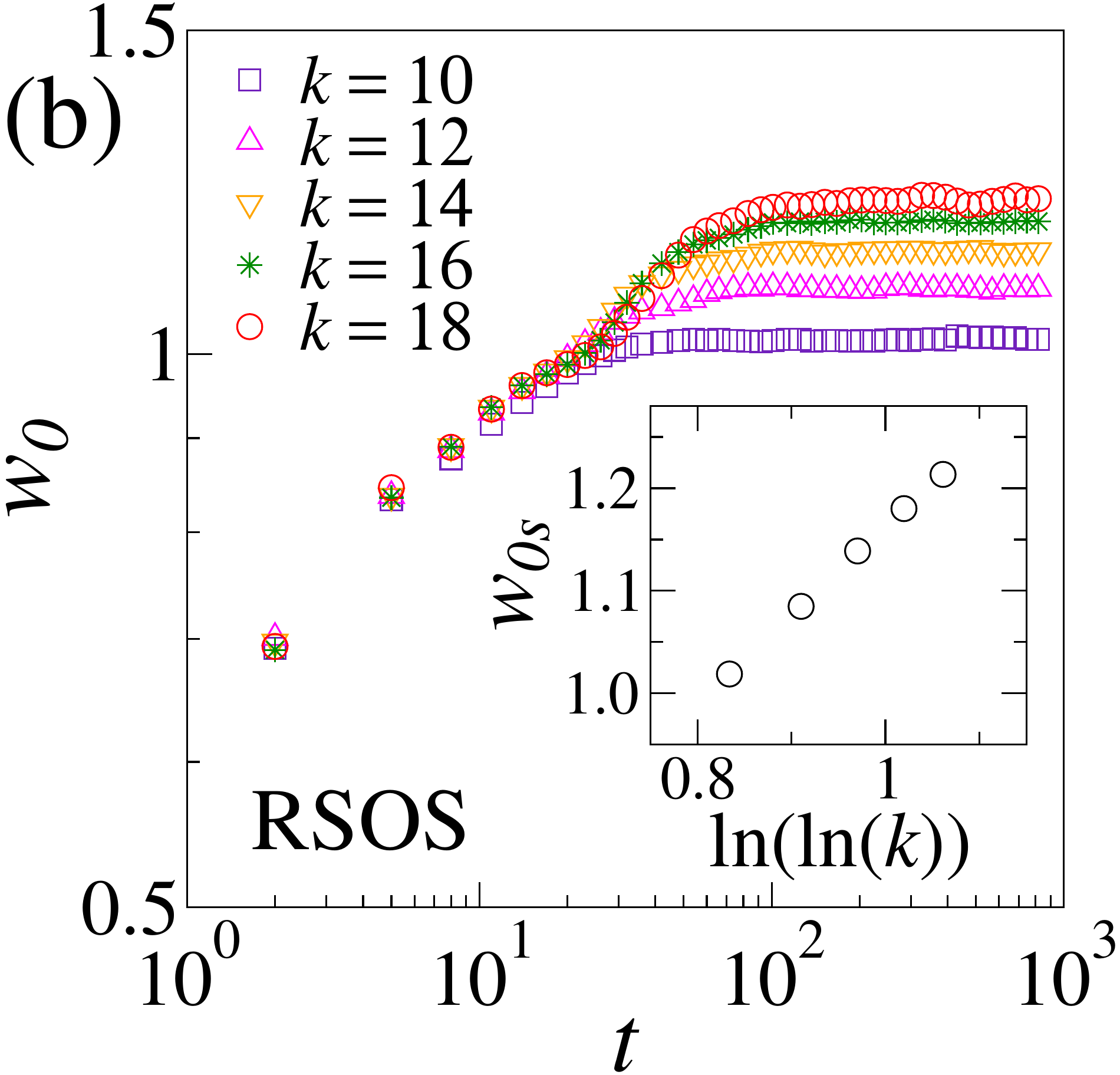}
 \includegraphics[width=4.25cm]{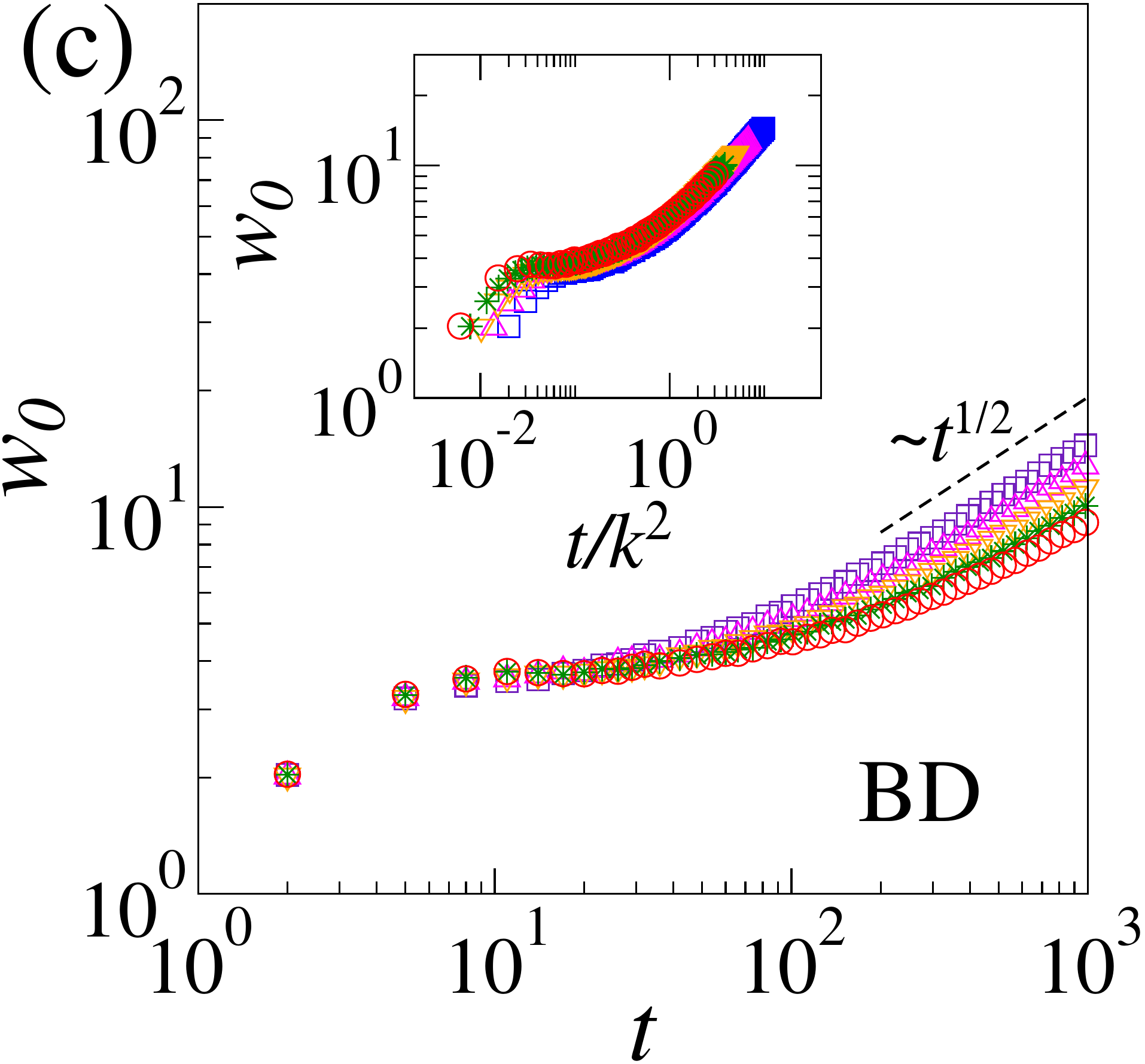}
 \includegraphics[width=4.25cm]{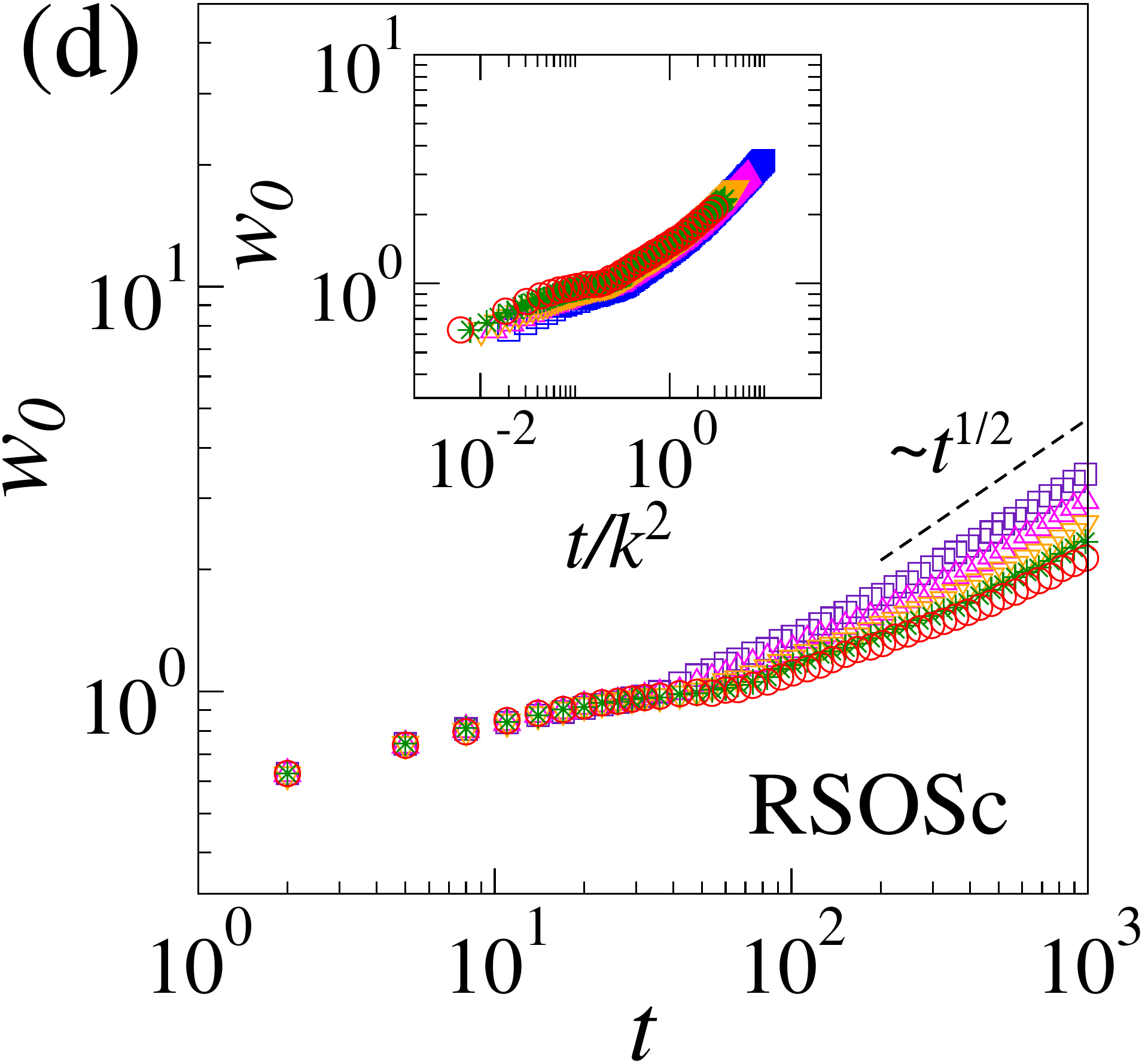}
 \caption{Log-log plots of the 1-pt height fluctuations $w_0$ versus time $t$ for the a) RDSR, b) RSOS, c) BD and d) RSOSc models. Panel (a) shows data for deposition on hypercubic substrates of lateral size $L$ and dimensions $d=3$ (with $L \in [64,512]$), $d=4$ (with $L \in [16,128]$) and $d=5$ (with $L \in [8,48]$), as well as on CTs with $q=3$ and $k\in [8,12]$, whose increase is indicated by the arrow. Results in (b)-(d) are for CTs with $q=3$ and the same $k$'s indicated in the legend of (b). The saturation values of $w_0$ in (b) are displayed in the insert against $\ln(\ln(k))$, in linear-linear scale. The insertions in (c) and (d) present the same data from the main plots with the time rescaled by $k^2$. The dashed lines in (c) and (d) have the indicated slopes.}
 \label{fig3}
\end{figure}

In view of this and bearing in mind the interest in assessing the behavior of growth models on the BL, hereafter I will investigate the height fluctuations only at the central site ($i=0$) of the CT. Namely, I will analyze the standard deviation of the height distribution $P(h_0)$, being $w_0 = \sqrt{\left\langle h_0^2 \right\rangle - \left\langle h_0 \right\rangle^2}$, where $\left\langle \cdot \right\rangle$ denotes average over different surfaces, at a given time. I remark that, during the growth regime, the height at a single point of a growing surface is expected to evolve as $h \simeq v_{\infty} t + (\Gamma t)^{\beta} \chi + \ldots$ \cite{Prahofer2000}, for dimensions $d < d_u$. At $d=d_u$ this changes to $h \simeq v_{\infty} t + \sqrt{\Gamma \ln(t)} \chi + \ldots$, as recently demonstrated for the EW class in $d=2$ \cite{Ismael19}. In these \textit{ans\"atze}, which are expected to be valid regardless the surface is flat or curved, $\chi$ is a random variable fluctuating according to universal height distributions, $\Gamma$ is a model-dependent parameter and ``$\ldots$'' indicates possible non-universal corrections. Note that this yields the scaling of the 1-pt height fluctuations as $w_0 \sim t^{\beta}$ for $d < d_u$  and $w_0 \sim \sqrt{\ln (t)}$ when $d=d_u$. Above $d_u$, notwithstanding, the noise and possible nonlinearities are not capable of yielding a building up of height fluctuations, meaning that the surfaces shall be smooth \cite{barabasi}. Namely, when $d>d_u$, $w_0$ shall not scale in time, not even logarithmically, presenting at most small variations at short times, so that the \textit{ans\"atze} above are not valid. This is indeed confirmed in Fig. \ref{fig3}a, where the temporal evolution of $w_0$ is displayed for the RDSR model simulated on hypercubic substrates of dimensions $d=3$, $4$ and $5$, with periodic boundary conditions. For the sake of simplicity, in such simulations the freshly deposited particles were allowed to diffuse only to their NN sites. Moreover, since the surfaces are flat in this case, $w_0$ was measured considering all substrate sites. A noticeable, but quite slow temporal variation is observed only for $d=3$, while in higher dimensions one has $w_0 \sim const.$ already at very short times. A similar behavior is expected also for $d = \infty$.

The 1-pt height fluctuations for the RDSR model at the central site of CTs with different generations $k$ are also displayed in Fig. \ref{fig3}a. In this case, one sees that $w_0$ has a more complex variation, initially increasing, at very short times, then passing through a maximum, which seems to become a plateau for very large (and unfeasible) $k$'s, and finally decreasing towards its asymptotic value. This behavior is probably related to that observed in Fig. \ref{fig2}e, with the relaxation in surface shape yielding a slight smoothening in the 1-pt fluctuations. The key point to note here, however, is that $w_0$ is always very small, as well as its variation. For instance, the differences between the maxima and the saturation values in Fig. \ref{fig3}a are $\Delta w_0 \lesssim 0.03$. Moreover, $w_0$ does not scale in time (in contrast, it does even decrease) and its saturation values display only a very small variation with $k$, for small CTs. All these results strongly indicate that the height fluctuations at the central site of the CT will remain very small ($w_0 \sim 0.6$) in the asymptotic limit (when the CT size and the time go to infinity), as indeed expected for EW systems at $d = \infty$.

Figure \ref{fig3}b shows the temporal evolution of $w_0$ for the RSOS model, for which a mild increase is observed at very short times, followed by a saturation in small values $w_{0s} \sim 1$. Although these saturation values increase with $k$, such variation is quite slow. For instance, by comparing $w_{0s}$ for $k=10$ and $k=18$ one finds that the fluctuations increase by less than 20\% (being $\Delta w_{0s} \approx 0.19$), while the number of substrate sites increases by more than 25000\%. Another evidence of the very small dependence of $w_{0s}$ with $k$ is given in the insertion of Fig. \ref{fig3}b, which shows $w_{0s}$ versus $\ln(\ln(k))$ and an apparent linear behavior is found. A close look at the data for the larger $k$'s reveals however that $w_{0s}$ increases even slower than $\ln(\ln(k))$. These results strongly suggest that the surfaces of the RSOS model will be smooth even at the center of infinite CTs (i.e., on the BL), similarly to those of the RDSR model.

The temporal variation of $w_0$ for the BD model is depicted in Fig. \ref{fig3}c. In this case, after an increase at very short times, the 1-pt fluctuations tends to form a plateau at $w_0 \sim 3.6$, but then it crosses over to a regime where $w_0 \sim t^{1/2}$. The very same behavior is found for the RSOSc model, as shows Fig. \ref{fig3}d, where $w_0$ tends to form a plateau at $w_0 \sim 1$ before the onset of the asymptotic scaling $w_0 \sim t^{1/2}$. I remark that this apparent random growth (with $w_0 \sim t^{1/2}$) is actually the steady state regime, where, at first sight, $w_0$ should saturate, as it indeed does in the RDSR and RSOS models. The saturation is not observed in the BD and RSOSc models, due to the fluctuations in their (global) average heights $\bar{h}(t)$. I stress that this is a general feature of the 1-pt height fluctuations, having no relation with the CT. 

\begin{figure}[t]
 \includegraphics[width=4.25cm]{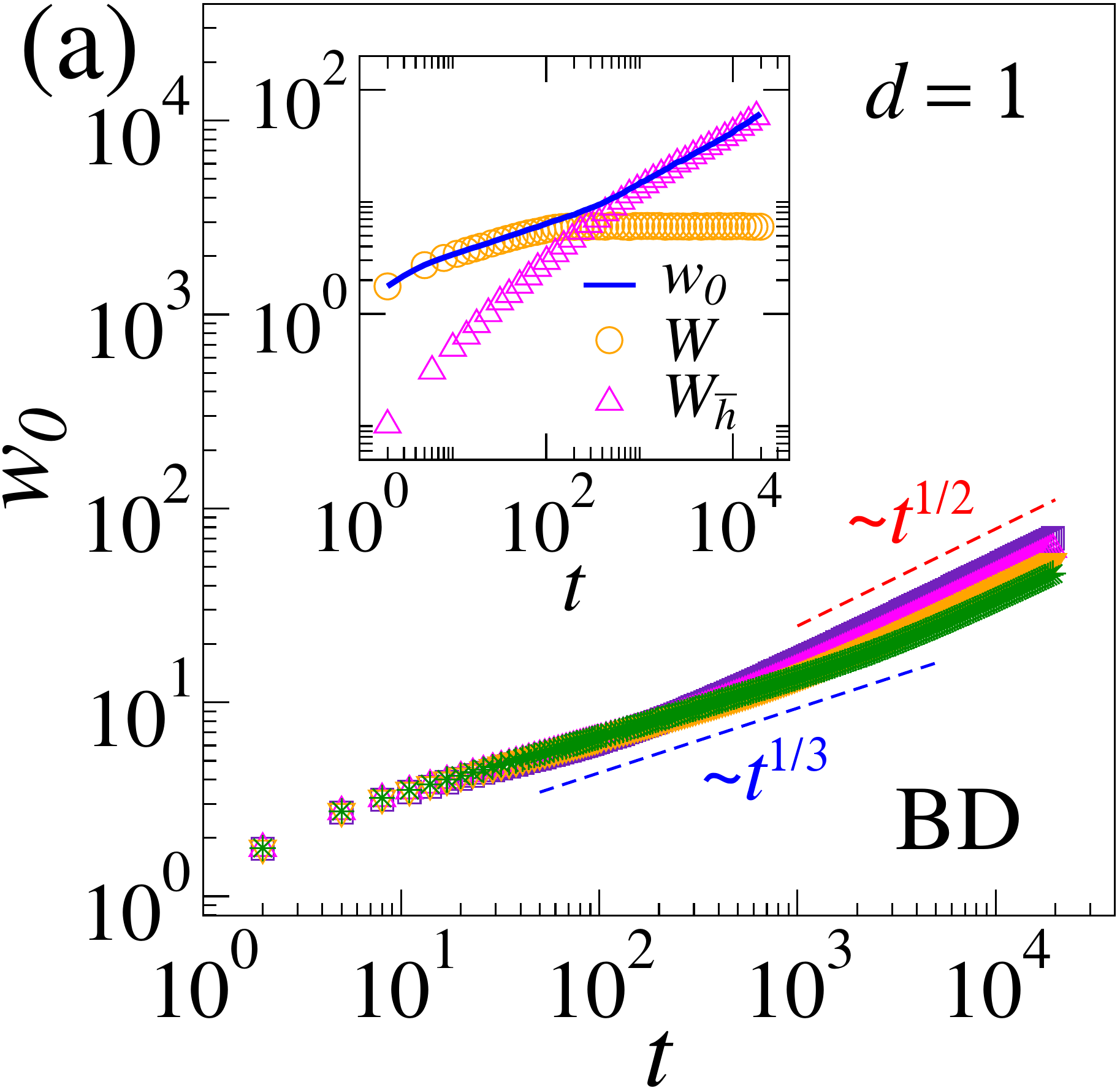}
 \includegraphics[width=4.25cm]{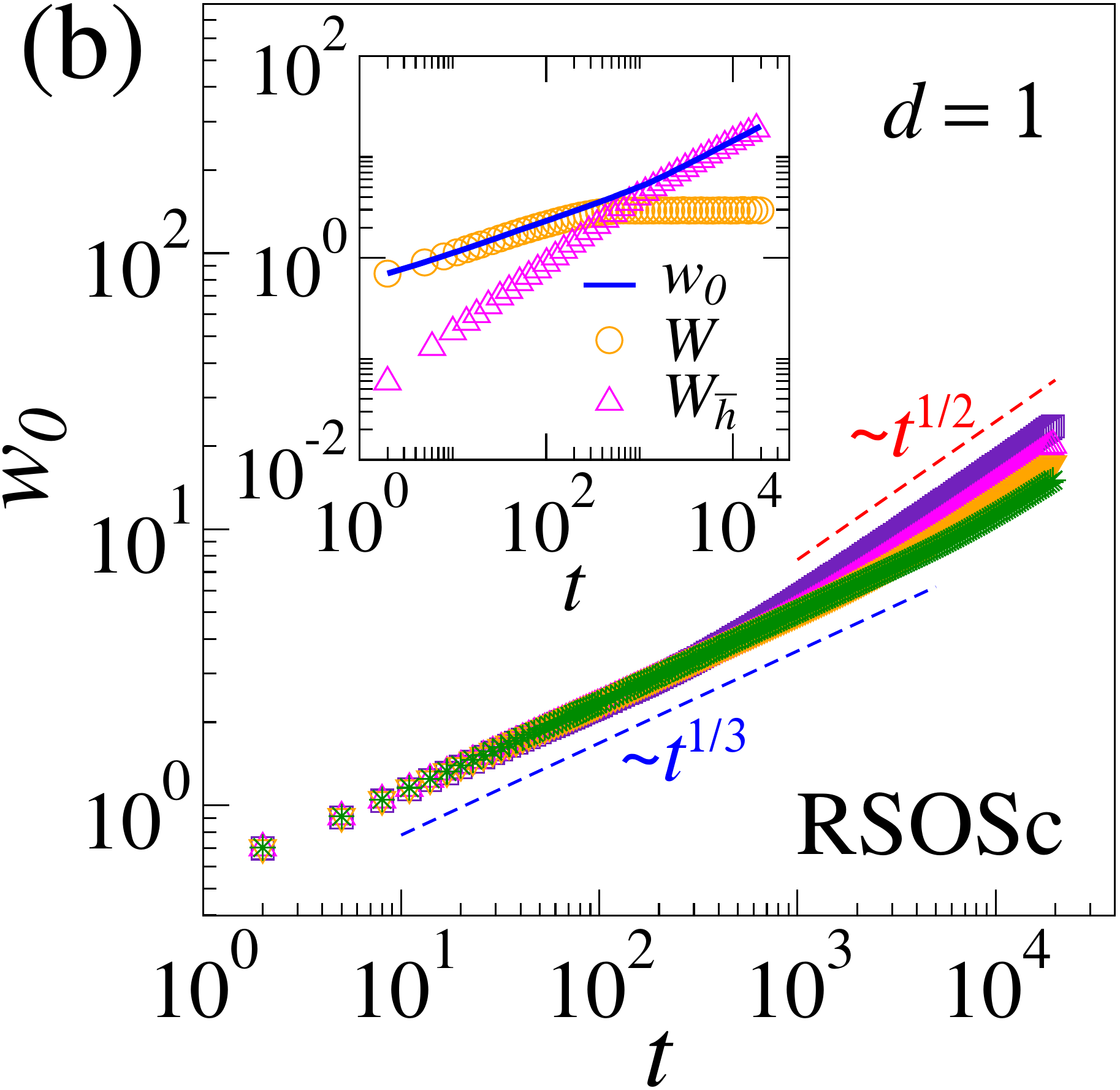}
 \caption{1-pt height fluctuations $w_0$ versus time $t$ for the a) BD and b) RSOSc models simulated on one-dimensional substrates of sizes $L\in [64,512]$. The insertions present the temporal evolution of $w_0$, $W$ and $W_{\bar{h}}$ for the same models with $L=128$. The dashed lines have the indicated slopes.}
 \label{fig4}
\end{figure}

In fact, if one grows $M$ samples for a given growth model on regular substrates [initially flat, with $L^d$ sites in $d$-dimensions and periodic boundary conditions (PBCs)], their surfaces will be flat (on average) and the squared global roughness can be defined as $W^2 = \langle \overline{h^2} - \bar{h}^2 \rangle$, where $\bar{\cdot}$ and $\left\langle \cdot \right\rangle$ respectively denote averages over the $L^d$ sites of a given surface and over the $M$ samples. Thanks to the spatial translation-invariance, the 1-pt height fluctuations can be spatially averaged, so that $w_0^2 = \langle \overline{h^2} \rangle - \overline{\langle h \rangle^2}$. In addition, this invariance implies also that $\overline{\langle h \rangle^2} = \langle \bar{h} \rangle^2$, which allow us to write $W^2= \langle \overline{h^2} \rangle - \overline{\langle h \rangle^2} - (\langle \bar{h}^2 \rangle - \langle \bar{h} \rangle^2)$. Then, it immediately follows that $w_0^2 = W^2 + W_{\bar{h}}^2$, where $W_{\bar{h}}^2 = \left\langle \bar{h}^2 \right\rangle - \left\langle \bar{h} \right\rangle^2$ is the variance of the \textit{average height} distribution $P_{av}(\bar{h})$. Namely, $w_0^2$ is given by the sum of fluctuations about $\bar{h}$ and the ones in $\bar{h}$. In models where $\bar{h}$ is deterministic, as is the case in the RDSR and RSOS models (as defined here), $W_{\bar{h}}=0$ and, thus, $w_0 = W$. Thereby, the 1-pt fluctuations behave exactly as the global roughness, saturating when the system becomes completely correlated. I have indeed confirmed this for these models simulated on 1D and 2D substrates with PBCs. On the other hand, in the BD, RSOSc and several other models (including non-KPZ ones) for which $\bar{h}$ fluctuates, the equivalence between $w_0$ and $W$ is lost, due to $W_{\bar{h}} \neq 0$. These quantities (\textit{vs} time) are compared in the insertions of Figs. \ref{fig4}a and \ref{fig4}b, for the BD and RSOSc models deposited on 1D substrates of size $L=128$ with PBCs. At short times, $W_{\bar{h}} \ll W$ and then $w_0 \sim W$, whereas at long times $W$ saturates, yielding $w_0 = \sqrt{const + W_{\bar{h}}^2}$ and asymptotically $w_0 \sim W_{\bar{h}}$. In turn, it is widely known that $W_{\bar{h}} \sim t^{1/2}$ in the stationary regime of 1D KPZ and EW systems \cite{DerridaLDF,*DerridaLDF2,*LeeLDF,*Appert,*LeeLDF2,*healyEW} and this seems to be a general behavior. In fact, I have verified this also for the BD and RSOSc models in $d=2$ and, importantly here, on the CT. Hence, in this class of models the steady state regime is characterized by $w_0 \sim t^{1/2}$, independently of $d$. 

In curved surfaces where the spatial translation-invariance is lost, as is the case here on the CT, it is quite expected that $w_0$ (measured at a single point) will display the same behavior just discussed for flat surfaces. Namely, at short times it shall scale according to the fluctuations about $\bar{h}$, while at long times it will be ruled by the fluctuations in $\bar{h}$. This is fully consistent with (and certainly explains) the different asymptotic behaviors found in Fig. \ref{fig3} for the fluctuations at the center of the CT.

Although the steady state scaling of $w_0$ does not allow us to draw conclusions about $d_u$ for the BD and RSOSc models, the temporal variation of $w_0$ during the growth regime allow it. For instance, one sees in Fig. \ref{fig4} a good agreement with $w_0 \sim t^{1/3}$ at short times, as expected for 1D KPZ systems, while Figs. \ref{fig3}c and \ref{fig3}d reveal a quite different behavior for the same models on the CT. Substantially, since the crossover to the stationary regime occurs in a time, $t_c$, which increases with $k$, this strongly indicates that on infinite CTs $w_0$ will have only plateaus at $w_0 \sim 3.6$ in BD and $w_0 \sim 1$ in the RSOSc case, during the growth regime. So, once again, these KPZ models shall display a behavior similar to that for the RDSR model at $d=\infty$. Interestingly, I find that $t_c \sim k^2$ for these models, as demonstrated by the striking data collapse in the insertions of Figs. \ref{fig3}c and \ref{fig3}d, which show the same data in the main plots with the time rescaled by $k^2$. Since the chemical distance between the central site of the CT and any site at its boundary is $k$, it is tempting to assume that $t_c \sim k^z$, with $z=\alpha/\beta$ being the dynamic exponent. In such a case, the exponent $z = 2$ (which is a hallmark of the EW class) would provide another compelling evidence that the KPZ models are indeed behaving as expected for $d \geqslant d_u$. It turns out however that the definition of distances in the CT is a difficult task and the use of the chemical distance to measure correlations on this tree has been a source of confusion in literature (see, e.g., Ref. \cite{JurgenCTEuc} for a discussion on this issue).

Even though only results for CTs with coordination $q=3$ were presented above, I have verified that everything is quite similar for small CTs with $q=4$ and $q=5$. The noteworthy differences are that, by increasing $q$, the surfaces become smoother (i.e., in most cases $w_0$ decreases with $q$, for a given time and $k$) and the boundary effects are enhanced, as expected, leading, e.g., $\Delta \langle h \rangle$ to increase.

In summary, I have demonstrated that, as a consequence of boundary effects, growth models develop curved surfaces when deposited on CTs. Importantly, at the center of the CT, which one may see as an approximation for the Bethe lattice, the height fluctuations present only mild variations in time, during the growth regime, for all investigated models. Moreover, their saturation values practically do not increase with the CT size (in the models for which $w_0$ saturates). This provides strong evidence that the surfaces of both EW and KPZ models are asymptotically smooth, consistently with the behavior expected for these growing systems at $d=\infty$. As an aside, I notice that another common way of studying models, in general, in the infinite-dimensional limit is by defining them on a complete graph. Although it is not so easy to predict what happens in the BD model if one simulates it considering complete graphs as substrates, for the RDSR and RSOS/RSOSc models it is trivial to see that a perfect layer-by-layer growth will occur, since all substrate sites will be nearest neighbors. Therefore, their surfaces will be almost perfectly smooth, once again, as expected for these growing systems at $d =\infty$. Hence, if one assumes that these models are still belonging to their respective universality classes when defined on these effectively infinite-dimensional lattices, one may conclude that, at least at $d = \infty$, KPZ systems behaves as the EW ones. Whether this is valid also for finite, but high dimensions (in case of a finite $d_u$ for the KPZ class) it is still an open question.

\acknowledgments

The author acknowledges financial support from CNPq and FAPEMIG (Brazilian agencies). I thank J\"urgen F. Stilck for a critical reading of this manuscript and helpful discussions.

\bibliographystyle{eplbib}
\bibliography{bibBetheGrowth}

\end{document}